\documentclass[fleqn,twoside]{article}
\usepackage{espcrc2}

\usepackage{epsfig,graphics,graphicx}

\newcommand{\AmS}{{\protect\the\textfont2
  A\kern-.1667em\lower.5ex\hbox{M}\kern-.125emS}}
\newcommand{\U}{\mathrm{U}}
\newcommand{\Th}{\mathrm{Th}}
\newcommand{\K}{\mathrm{K}}
\title{Geo-Neutrinos: a short review}

\author{Gianni Fiorentini\address{Dipartimento di Fisica dell'Universit\`a  di Ferrara
         and INFN Sezione di Ferrara, I-44100 Ferrara, Italy},
         Marcello Lissia\address{INFN Sezione di Cagliari and Dipartimento di Fisica
         dell'Universit\`a di Cagliari, I-09042 Monserrato (CA),
         Italy},
         Fabio Mantovani\address{Dipartimento di Scienze della Terra,
         Universit\`a  di Siena, I-53100 Siena,  Centro di GeoTecnologie
         CGT, I-52027 San Giovanni Valdarno and INFN Sezione di Ferrara, I-44100 Ferrara, Italy}
         and Riccardo Vannucci\address{Dipartimento di Scienze della Terra,
         Universit\`a  di Pavia, I-27100 Pavia, Italy}}

\begin{document}

\begin{abstract}
Geo-neutrino detection will determine the amount of long lived
radioactive elements within our planet and fix the debated
radiogenic contribution to the terrestrial heat. In addition, it
will provide a direct test of the Bulk Silicate Earth model, a
fundamental cosmochemical paradigm about the origin  of the Earth.
Unorthodox models of  Earth's core (including the presence of
potassium or the possibility of a  giant reactor) can also be
checked. This short review presents  status and  prospects of the
field.
 \vspace{1pc}
\end{abstract}

\maketitle

\section{PROBES OF THE EARTH'S INTERIOR}

The deepest hole that has ever been dug is about 12~km deep, a
mere dent in planetary terms. Geochemists analyze samples from the
Earth's crust and from  the top of the  mantle. Seismology can
reconstruct the density profile throughout all Earth, but not its
composition. In this respect, our planet is mainly unexplored.

Geo-neutrinos, the antineutrinos from the progenies of  U, Th and
$^{40}$K decays in the Earth, bring to the surface information
from the whole planet, concerning its content of radioactive
elements. Their detection can shed light on the sources of the
terrestrial heat flow, on the present composition and on the
origin of the Earth.

They  represent a new probe of our planet, which is becoming
practical as a consequence of two fundamental advances that
occurred in the last few years:  a) development of extremely low
background neutrino detectors and b) progress on understanding
neutrino propagation.

Geo-neutrino properties  are summarized in
Table~\ref{table:nuproperties}, where the last two columns present
the heat and anti-neutrino production rates per unit mass and
natural isotopic composition.

\begin{table*}[htb]
\caption{The main properties of geo-neutrinos.}
\label{table:nuproperties}
\newcommand{\dg}{\hphantom{$0$}}
\newcommand{\cc}[1]{\multicolumn{1}{c}{#1}}
\renewcommand{\arraystretch}{1.2} 
\begin{tabular}{llllll}
\hline
Decay           & \cc{$Q$} & \cc{$\tau_{1/2}$}
& \cc{$E_{\mathrm{max}}$} & \cc{$\epsilon_{H}$} & \cc{$\epsilon_{\bar{\nu}}$}\\
      &  \cc{[MeV]} & \cc{[$10^9$~yr]} & \cc{[MeV]} &
      \cc{[W/Kg]} & \cc{[kg$^{-1}$s$^{-1}$]} \\
\hline $^{238}\mathrm{U\phantom{h}} \to {}^{206}\mathrm{Pb} + 8\,
{}^{4}\mathrm{He} + 6 e + 6 \bar{\nu}$
&   51.7 & \dg4.47& 3.26 & $0.95\times 10^{-4}$ & $7.41\times 10^{7}$ \\
$^{232}\mathrm{Th} \to {}^{208}\mathrm{Pb} + 6\, {}^{4}\mathrm{He}
+ 4 e + 4 \bar{\nu}$
&   42.7 &   14.0 & 2.25 & $0.27\times 10^{-4}$ & $1.63\times 10^{7}$ \\
$^{\phantom{0}40}\mathrm{K\phantom{h}} \to
{}^{\phantom{0}40}\mathrm{Ca} + e + \bar{\nu}$
& \dg1.32& \dg1.28& 1.31 & $0.36\times 10^{-8}$ & $2.69\times 10^{4}$ \\
\hline
\end{tabular}\\[2pt]
\end{table*}

For each element there is a strict connection between the
geo-neutrino luminosity $L$ (anti-neutrinos produced in the Earth
per unit time), the radiogenic heat production rate $H_R$ and the
mass $m$ of that  element in the Earth:
\begin{equation}\label{eq:luminosity}
L= 7.4\times m(\mathrm{U}) + 1.6\times m(\mathrm{Th}) + 27\times
m(^{40}\mathrm{K})
\end{equation}
\begin{equation}\label{eq:heat}
H_R= 9.5\times m(\mathrm{U}) + 2.7\times m(\mathrm{Th}) +
3.6\times m(^{40}\mathrm{K})
\end{equation}
where units are $10^{24}$~s$^{-1}$, $10^{12}$~W and $10^{17}$~kg,
respectively.

Geo-neutrinos originating  from different elements can be
distinguished due to their different energy spectra, e.g.,
geo-neutrinos with $E >2.25$~MeV are produced only in the Uranium
chain. Geo-neutrinos from U and Th (not those from $^{40}$K) are
above threshold for the classical anti-neutrino detection
reaction, the inverse beta on free protons:
\begin{equation}\label{eq:inversebeta}
\bar{\nu} + p \to e^{+} + n - 1.8\mathrm{\ MeV} \quad .
\end{equation}
Anti-neutrinos from the Earth are not obscured by solar neutrinos,
which cannot yield reaction~(\ref{eq:inversebeta}).

In this short review we shall concentrate  on geo-neutrinos from
Uranium, which are closer to experimental detection, and on the
predictions for Kamioka site hosting KamLAND~\cite{Eguchi:2002dm},
the only detector which is presently operational.

\section{A BIT OF HISTORY}

Geo-neutrinos  were introduced by Eder~\cite{Eder} in the sixties
and Marx~\cite{Marx}  soon realized their relevance. In the
eighties Krauss et al. discussed their potential as probes of the
Earth's interior in an extensive publication~\cite{Krauss:zn}. In
the nineties the first paper on a geophysical journal was
published by Kobayashi et al.~\cite{Kobayashi}.  In 1998, Raghavan
et al.~\cite{Raghavan:1997gw} and Rotschild et
al.~\cite{Rothschild:1997dd} pointed out that KamLAND and Borexino
should be capable of geo-neutrino detection.

In the last two years more papers appeared than in the preceding
millennium: in a series of papers Fiorentini et
al.~\cite{Fiorentini:2002bp,Fiorentini:2003ww,Mantovani:2003yd}
discussed the potential of geo-neutrinos for determining the
radiogenic contribution to the terrestrial heat flow and for
discriminating among different models of  Earth's composition and
origin.

The indication of geo-neutrinos in  the first data release from
KamLAND~\cite{Eguchi:2002dm} was a most important point which
stimulated several investigations
\cite{Nunokawa:2003dd,Mitsui:2003fm,Miramonti:2003hw,%
Domogatski:2004gs,Domogatski:2004zz,McKeown:2004yq,Fields:2004tf,%
Rusov:2003sx,Fogli:2004vb}.

\section{ENERGETICS OF THE EARTH AND THE  MISSING HEAT SOURCE
MYSTERY}

There is a tiny flux of heat coming from the Earth. It  depends on
the site and  is generally of the order of 60~mW/m$^2$.  By
suitably integrating over the Earth surface one obtains  a total
flow $H_E$ in the range 30-45 TW, the equivalent of some $10^4$
nuclear plants.  A frequently quoted estimate is  $H_E =(44\pm 1
)$~TW~\cite{Pollack:1993}, where the statistical error does not
account for the systematic uncertainties (in particular concerning
the contributions of the oceanic crust). For sure, heat released
from radiogenic elements is important, however its role  is not
understood at a quantitative level.

Verhoogen in 1980~\cite{Verhoogen:1980} makes the following
summary : ``\ldots What emerges from this morass of fragmentary
and uncertain data is that radioactivity itself could possibly
account for at least 60 per cent if not 100 per cent of the
Earth's heat output \ldots If one adds the greater rate of
radiogenic heat production in the past, possible release of
gravitational energy (original heat, separation of the core
\ldots) tidal friction \ldots and possible meteoritic impact
\ldots the total supply of energy may seem embarrassingly large''.

In a recent paper with the same title as this paragraph,
Anderson~\cite{Anderson} has  a more cautious approach: ``Global
heat flow estimates range from 30 to 44 TW \ldots Estimates of the
radiogenic contribution \ldots based on cosmochemical
considerations, vary from 19 to 31 TW. Thus, there is either a
good balance between current input and output, as was once
believed, \ldots or there is a serious missing heat source
problem, up to a deficit of 25 TW''.

We remark that the radiogenic component is  essentially based on
cosmo-chemical considerations and that a direct determination, as
offered by geo-neutrino detection, is important.

\section{U, Th AND K IN THE EARTH: HOW MUCH AND WHERE?}

Earth global composition is generally estimated from that of CI
chondritic meteorites by using geochemical arguments which account
for loss and fractionation during planet formation. Along these
lines the Bulk Silicate Earth (BSE) model is built, which
describes the ``primitive mantle'', i.e.,  the outer portion of
the Earth after core separation and before  the differentiation
between crust and mantle. The model is believed to describe the
present crust plus mantle system. It  provides  the total amounts
of U, Th and K in the Earth, as these lithophile elements should
be absent in the core. Estimates from different
authors~\cite{Mcdonough:2003} are concordant within 10-15\%. From
the mass, the present radiogenic heat production rate and neutrino
luminosity can be immediately calculated by means of Eqs.
(\ref{eq:luminosity}) and (\ref{eq:heat}) and are are shown in the
following Table~\ref{table:UThKinBSE}.
\begin{table}[htb]
\caption{U, Th and K according to BSE} \label{table:UThKinBSE}
\newcommand{\dg}{\hphantom{$0$}}
\newcommand{\cc}[1]{\multicolumn{1}{c}{#1}}
\renewcommand{\arraystretch}{1.2} 
\begin{tabular}{llll}
\hline
       & \cc{$m$}            & \cc{$H_R$} & \cc{$L_{\nu}$} \\
       & \cc{[$10^{17}$~kg]} & \cc{[$10^{12}$~W]}
       & \cc{[$10^{24}$ s$^{-1}$]} \\
\hline
U        &  0.8 & 7.6 & \dg5.9 \\
Th       &  3.1 & 8.5 & \dg5.0 \\
$^{40}$K &  0.8 & 3.3 &   21.6 \\
\hline
\end{tabular}\\[2pt]
\end{table}

The BSE is a fundamental  geochemical paradigm. It is  consistent
with most  observations, which however regard  the crust and the
uppermost portion of the mantle only. Its prediction for the
present radiogenic production is 19 TW.

Concerning the distribution of radiogenic elements, estimates  for
Uranium in  the (continental) crust based on observational data
are in the range:
\begin{equation}
m_C(\mathrm{U}) = (0.3 - 0.4) 10^{17} \mathrm{ kg} \quad .
\end{equation}
The crust --- really a tiny envelope --- should thus contain about
one half of Uranium in the Earth.

For the mantle, observational data are scarce and restricted to
the uppermost part, so the best estimate for its Uranium content
$m_M(\mathrm{U})$ is obtained by subtracting the crust
contribution to the BSE estimate:
\begin{equation}
  m_M(\mathrm{U})= m_{\mathrm{BSE}}(\mathrm{U})- m_C(\mathrm{U})
  \quad .
\end{equation}

Compositionally, geochemists prefer a two-layered mantle,  the
lower part being closer to the primitive composition (Uranium mass
abundance  $a(\U)=20$~ppb), the upper part being  impoverished in
these elements, $a(\U)=(5-8)$~ppb. On the other hand,
seismological evidence points toward a fully mixed and thus
globally homogeneous mantle.

Similar considerations hold for Thorium and Potassium, the
relative mass abundance with respect to Uranium being globally
estimated as  $a(\Th) : a(\U) : a(\K) = 4 : 1 : 10,000$ .

 Geochemical arguments
are against the presence of radioactive elements in the
(completely unexplored) core, as discussed by McDonough  in an
excellent review  of compositional models of the
Earth~\cite{Mcdonough:2003}.

For a comparison, let us summarize some --- less orthodox or even
heretical --- alternatives to the canonical BSE model:

a) it is conceivable that the original material from which the
Earth formed is not wholly the same as inferred from
CI-chondrites. A model with initial composition as that of
enstatite chondrites could account for a present production of
some 30 TW~\cite{Javoy:1999,Hofmeister:2004}.

b) A model where the BSE abundances of U, Th and K  are
proportionally rescaled by a a factor of 2.3 cannot be excluded by
the observational data, if one assumes that the missing radiogenic
material is hidden below the upper mantle. This model gives a
present radiogenic heat production of  44 TW, the maximum which
can be tolerated by Earth energetics since  it takes time to bring
heat to the surface and more heat was produced in the past.

c) Several authors have been considering the possibility that a
large amount of Potassium  is sequestered into the Earth's core,
where it provides the light element to account for the right core
density, the energy source for driving the terrestrial dynamo and
--- more generally --- an additional  contribution to Earth energy
budget.

d) Herndon~\cite{Herndon} has  proposed that a large drop of
Uranium has been collected at the center  of the Earth, forming  a
natural 3-6 TW breeder reactor, see also~\cite{Raghavan:2002eh}.
In this case nuclear fission should provide the energy source for
terrestrial magnetic field, a contribution to missing heat, and
the source of the anomalous $^3$He/$^4$He  flow  from Earth.

In summary, the BSE is a fundamental  geochemical paradigm
accounting for the radiogenic production of about 19 TW.  It is
consistent with most observations, which however regard  the crust
and the uppermost portion of the mantle only, most of the Earth
being unexplored. It should be tested.

\section{FROM LUMINOSITY TO FLUX AND SIGNAL}
The goal with geo-neutrinos is the determination of the neutrino
luminosities $L$ produced in the Earth (for each element), which
immediately give the amounts of radioactive material in the
Earth's interior.

What is measured  is the angle integrated flux
$\Phi_{\mathrm{ar}}$ of $\bar{\nu}_e$ arriving at the detector
position. An order of magnitude estimate is immediately obtained
from:
\begin{equation}\label{eq:fluxestimate}
    \Phi \approx \frac{\langle P_{ee}\rangle L}{4\pi R^2}
    \quad ,
\end{equation}
where  $\langle P_{ee}\rangle =0.59$  is the average survival
probability and $R$ is the Earth's radius. This gives antineutrino
fluxes of order $10^{6}$~cm$^{-2}$s$^{-1}$, comparable to that of
$^8$B neutrinos from the Sun.  From the cross section for reaction
(\ref{eq:inversebeta}) the reaction rates $S(\U)$ and $S(\Th)$ in
a detector containing $N_p$ free protons are:
\begin{eqnarray}
  S(\U) &=& 13.2\frac{\Phi_{\mathrm{ar}}(\U)}{10^{6}\mathrm{
  cm}^{-2}\mathrm{s}^{-1}}\frac{N_p}{10^{32}} \mathrm{\ yr}^{-1}\\
  S(\Th) &=&4.0\frac{\Phi_{\mathrm{ar}}(\Th)}{10^{6}\mathrm{
  cm}^{-2}\mathrm{s}^{-1}}\frac{N_p}{10^{32}}\mathrm{\ yr}^{-1}
  \quad .
\end{eqnarray}
This gives some tens of events per year in a kiloton detector.

For a  precise estimate of the flux as a function of the amount
$m$ of the parent element  in the Earth  one needs to know the
distribution of that  element  inside  the Earth. This involves
several steps, which we shall elucidate for Uranium geo-neutrinos:

i) For the world crust, one resorts to geological maps of the
Earth crust. A $2^{\circ}\times 2^{\circ}$ map, distinguishing
seven crust layers, has been used in Ref.~\cite{Mantovani:2003yd}.
Concerning element abundances, for each layer minimal and maximal
estimates present in the literature are adopted, so as to obtain a
range of acceptable fluxes. Depending on the adopted values, the
Uranium mass in the crust $m_C(\U)$ is in between $(0.3 - 0.4)
\times 10^{17}$~kg, the larger the mass, the bigger the signal.

ii) For Uranium in the mantle, one assigns to it a mass $m_M (\U)
= m(\U)-m_C(\U)$.  Generally, the minimal (maximal) contributed
flux is obtained by placing this Uranium as far (close) as
possible to the detector~\cite{Fiorentini:2004rj}. By assuming
spherical symmetry in the mantle and that the Uranium mass
abundance is a non decreasing function of depth the two cases
corresponds respectively  to: (a) placing Uranium in a thin layer
at the bottom and (b) distributing it with uniform abundance over
the mantle.

iii) This argument can be used again to combine the flux from
crust and mantle: for a fixed total $m$, the highest flux is
obtained by assigning to the crust as much as consistent  with
observational data ($m_C(\U)=0.4$) and putting  the rest
$m(\U)-m_C(\U)$ in the mantle with a uniform distribution.
Similarly the minimal flux is obtained for the minimal mass in the
crust ($m_C(\U)= 0.3$) and the rest in a thin layer at the bottom
of the mantle.

We remark that \emph{this argument, combining global mass balance
with geometry, is very powerful in constraining the range of
fluxes, which come out to be determined in a range of about
$\pm10\%$ for a fixed value of $m(\U)$}.

For a full exploitation of this information one needs a more
detailed geochemical and geophysical study of the region within a
few hundreds kilometers from the detector, where some half of the
signal is generated. The goal is to reduce the error on the
regional contribution to the level of the uncertainty on the rest
of the world. This has been recently
performed~\cite{Fiorentini:inpreparation}  for the region near the
KamLAND detector, which has been analyzed using geochemical
information on a  $0.25^{\circ}\times 0.25^{\circ}$ grid and a
detailed map of the crust depth.  The possible (minimal and
maximal) effects of the Pacific slab subducting beneath Japan are
considered  and the uncertainty arising from the debated
(continental or oceanic) nature of the crust below the Japan sea
is taken into account.

\begin{figure}[htb]
\epsfig{file=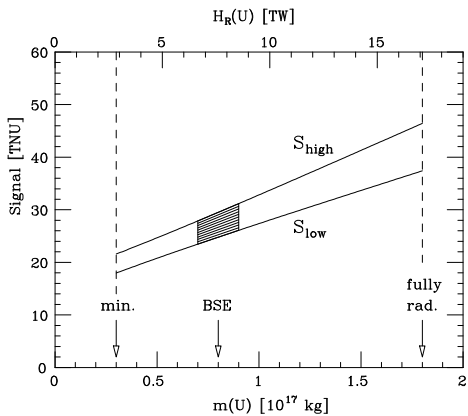,width=6cm,angle=90} \caption{The
predicted signal from Uranium geo-neutrinos at KamLAND.}
\label{fig:Kamlandsignalmass}
\end{figure}

The expected signal from Uranium at KamLAND is presented as a
function of the total uranium mass $m(\U)$ in
Fig.~\ref{fig:Kamlandsignalmass}~\cite{Fiorentini:inpreparation}.
The upper horizontal scale indicates the corresponding radiogenic
heat  production rate from Uranium. The signal is given  in
Terrestrial Neutrino Units:
\begin{equation}\label{eq:TNU}
    1 \mathrm{\ TNU} = 1 \mathrm{\ event} / (10^{32}\mathrm{\ protons} \cdot
    \mathrm{yr})
    \quad .
\end{equation}

The predicted signal as a function of $m(\U)$  is between the two
lines denoted  as $S_{\mathrm{low}}$ and $S_{\mathrm{high}}$.

Since the minimal amount of Uranium in the Earth is $0.3\times
10^{17}$~kg (corresponding to the minimal estimate in the crust
and a negligible amount in the mantle), we expect a signal of at
least 18~TNU.

The maximal amount of Uranium tolerated by Earth energetics,
$1.8\times 10^{17}$~kg, implies a signal not exceeding 46 TNU.

We remark that estimates by different authors for the Uranium mass
within the BSE are all between $(0.7-0.9)\times 10^{17}$~kg. This
translates into:
\begin{equation}\label{eq:fluxlimitsBSE}
    23 < S(\U) <31 \mathrm{\ TNU} \quad .
\end{equation}
\emph{The measurement of  geo-neutrinos  can  thus provide a
direct test of an important paradigm.}

\section{LOOKING FORWARD TO NEW DATA}

At the end of 2002, in the first data release equivalent  to an
exposure $0.11 \times 10^{32}$ proton $\cdot$ yr and 100\%
efficiency, KamLAND reported~\cite{Eguchi:2002dm} 4 events from
Uranium and 5 from Thorium from a total of 32 counts in the
geo-neutrino energy region ($E_{\mathrm{vis}} < 2.6$~MeV), after
subtracting 20 reactor events and 3 background  counts.
Statistical fluctuations imply that the ($1\sigma$) error is,  at
least, 5.7 counts. This means:
\begin{equation}\label{eq:totalsignal}
    S(\mathrm{U}+\mathrm{Th}) = (82\pm52)\mathrm{ TNU} \quad .
\end{equation}

The uncertainty is so large that the result is just an indication
of geo-neutrinos.

By now, KamLAND has accumulated a much larger statistics (see the
talk by G. Gratta) and the group is presently analyzing data which
might provide a definite geo-neutrino signal.

The vicinity of many nuclear-power reactors, which was essential
for the study of neutrino oscillations, is a major drawback for
measuring geo-neutrinos, the signal ratio being
$S_{\mathrm{rea}}/S_{\mathrm{geo}} = 5-10$.

\begin{figure}[htb]
\epsfig{file=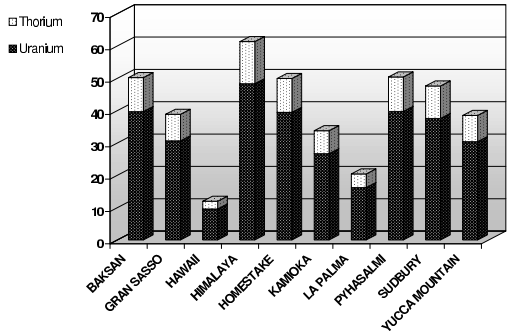,width=4.5cm,angle=-90}
\caption{Predicted signals, in TNU~\cite{Mantovani:2003yd}.}
\label{fig:signals}
\end{figure}

Several projects for geo-neutrino detection are being developed
(see Fig.~\ref{fig:signals} for the predicted signals at a few
locations). Borexino at Gran Sasso in Italy is expected to take
data in a few years. With respect to KamLAND, its smaller fiducial
mass can be compensated by the absence of nearby reactors
($S_{\mathrm{rea}}/S_{\mathrm{geo}} \approx 1$).

Mikaelyan et al. are proposing  a 1 Kton  scintillator detector in
Baksan, again very far from nuclear reactors.

A group at the Sudbury Neutrino Observatory in Canada is studying
the possibility of moving to liquid scintillator   after the
physics program with heavy water is completed. With  very low
reactor background and  in the middle of a well studied geological
environment it will have excellent opportunity for geo-neutrino
studies.

The LENA proposal envisages a 30 Kton liquid scintillator detector
at the Center for Underground Physics in the Pyh{\"a}salmi mine
(Finland). Due to the huge mass,  it should collect several
hundreds of events per year.

In conclusion, one can expect that within ten years the
geo-neutrino signal from Uranium and Thorium will be measured  at
a few points over the globe. This will fix the radiogenic
contribution of these elements to the terrestrial  heat  and will
provide a direct test of a fundamental paradigm on the origin and
the composition of our planet.

\section*{Acknowledgments}
We are grateful for useful comments and discussions to
C.~Bonadiman, L.~Carmignani, M.~Coltorti, K.~Inoue, E.~Lisi and
B.~Ricci. This work was partially supported by MIUR (Ministero
dell'Istruzione, dell'Universit\`a e della Ricerca) under
MIUR-PRIN-2003 project ``Theoretical Physics of the Nucleus and
the Many-Body Systems'' and MIUR-PRIN-2002 project ``Astroparticle
physics''.

\end{document}